\documentclass[12pt]{article}
\usepackage{amsmath}
\usepackage{amsfonts}
\usepackage{amssymb}
\usepackage{hyperref}
\hypersetup{colorlinks=true, linkcolor=blue, filecolor=magenta, urlcolor=cyan,}
\usepackage{graphicx}
\usepackage{tabularx}
\usepackage{ctable}
\usepackage{enumerate}
\usepackage{algorithm}
\usepackage{algpseudocode}
\usepackage[left=2.5cm,top=3cm,right=2.5cm,bottom=3cm,bindingoffset=0.5cm]{geometry}

\begin{document}
	
\title{\large\sc ESG driven pairs algorithm for sustainable trading: Analysis from the Indian market}
\normalsize
\author{\sc{Eeshaan Dutta} \thanks{Department of Mathematics, Indian Institute of Technology Guwahati, Guwahati-781039, India, e-mail: d.eeshaan@iitg.ac.in}
\and
\sc{Sarthak Diwan} \thanks{Department of Mathematics, Indian Institute of Technology Guwahati, Guwahati-781039, India, e-mail: s.diwan@iitg.ac.in}
\and 
\sc{Siddhartha P. Chakrabarty} \thanks{Department of Mathematics, Indian Institute of Technology Guwahati, Guwahati-781039, India, e-mail: pratim@iitg.ac.in, Phone: +91-361-2582606}
}

\date{}
\maketitle

\begin{abstract}

This paper proposes an algorithmic trading framework integrating Environmental, Social, and Governance (ESG) ratings with a pairs trading strategy. It addresses the demand for socially responsible investment solutions by developing a unique algorithm blending ESG data with methods for identifying co-integrated stocks. This allows selecting profitable pairs adhering to ESG principles. Further, it incorporates technical indicators for optimal trade execution within this sustainability framework. Extensive back-testing provides evidence of the model’s effectiveness, consistently generating positive returns exceeding conventional pairs trading strategies, while upholding ESG principles. This paves the way for a transformative approach to algorithmic trading, offering insights for investors, policymakers, and academics.

\end{abstract}

\section{Introduction}
\label{Sec_Introduction}

In a world that is increasingly interconnected and faced with challenges of climate change, the long term survival as well as success of a firm is contingent on factors beyond the realm of traditional metrics of financial performance. Accordingly, Environmental Social and Governance (ESG) considerations have emerged as a pivotal driver of sustainable performance of firms, not only from the perspective of investment decisions and risk management, but also from the viewpoint of consumer preferences, as well as regulatory practices, all of which, eventually has a bearing on the reputation and viability of a firms' operation \cite{wbcsd}. At the heart of ESG standing of a firm lies the encompassing of the environmental impact, which can be attributed to the firm, the stakeholders of the firm and the consequent practices of corporate governance. In this context, particular consideration is extended to environmental dimension, including carbon footprint, energy consumption and use, and the effort towards management of emission, reduction of waste and conservation of natural resources. 

In order to facilitate the tangible assessment of the ESG performance of a firm, in case of the environment component, one can adopt metrics such as per unit (of production) energy consumption, extent of green house gas (GHG) emission and level of recycling as well as conservation efforts \cite{wri}. The social dimension of ESG, emphasizes on the relationship that a firm shares with its stakeholders, both internal (ethical workplace practices, diversity and inclusion) and external (engagement with the local community and philanthropic outreach activities). Some of the metrics which can be adopted for this, include employee turnover, workplace safety, customer satisfaction and community engagement \cite{unpri}. Finally, on the criterion of good governance, the firm may be assessed via metrics like diversity indices, corporate governance ratings and performance on account of regulatory compliance.

Market-based tools like emissions trading schemes, carbon taxes, and carbon derivatives drive GHG emissions cuts. Emissions trading schemes work by capping emissions and issuing tradeable permits (which the firm can trade based on their emission levels). This enables reducing emission in a flexible and cost-effective manner \cite{EU_ETS}. Challenges with over-allocated permits, thereby leading to reduced prices can be addressed via mechanisms like the European Union Emission Trading Scheme (EU ETS) Market Stability Reserve, which reduces the surplus supply of permits \cite{Hellmich2021}.

The social cost of carbon, also called the carbon tax or carbon price, is the emissions tax per ton of CO2 \cite{Roncalli2022}. The tax level incentives shifts to low-carbon options across sectors. Examples of successful carbon tax initiatives exist in Nordic countries, with Sweden instituting a 1991 carbon tax covering about half its emissions. The tax design balances emissions cuts, competitiveness and social impacts. Revenues generated from carbon tax can also result in additional environmental and economic gains.

Additionally, carbon derivatives help firms hedge risks from carbon price fluctuations due to market-based tools. European companies covered by EU ETS use futures and option contracts on EU Allowances (EUAs) to provide price certainty over compliance periods \cite{ISDA}. As more countries implement carbon pricing, the carbon derivative markets is likely to experience growth, though transparency, regulation and liquidity challenges remain. Overall these flexible approaches mobilize private capital towards needed low-carbon investments, which is the genesis of this paper.

\section{Data Analysis and Asset Picking}
\label{Sec_Data_Analysis}

\begin{figure}[H]
\centering
\includegraphics[width=0.5\textwidth]{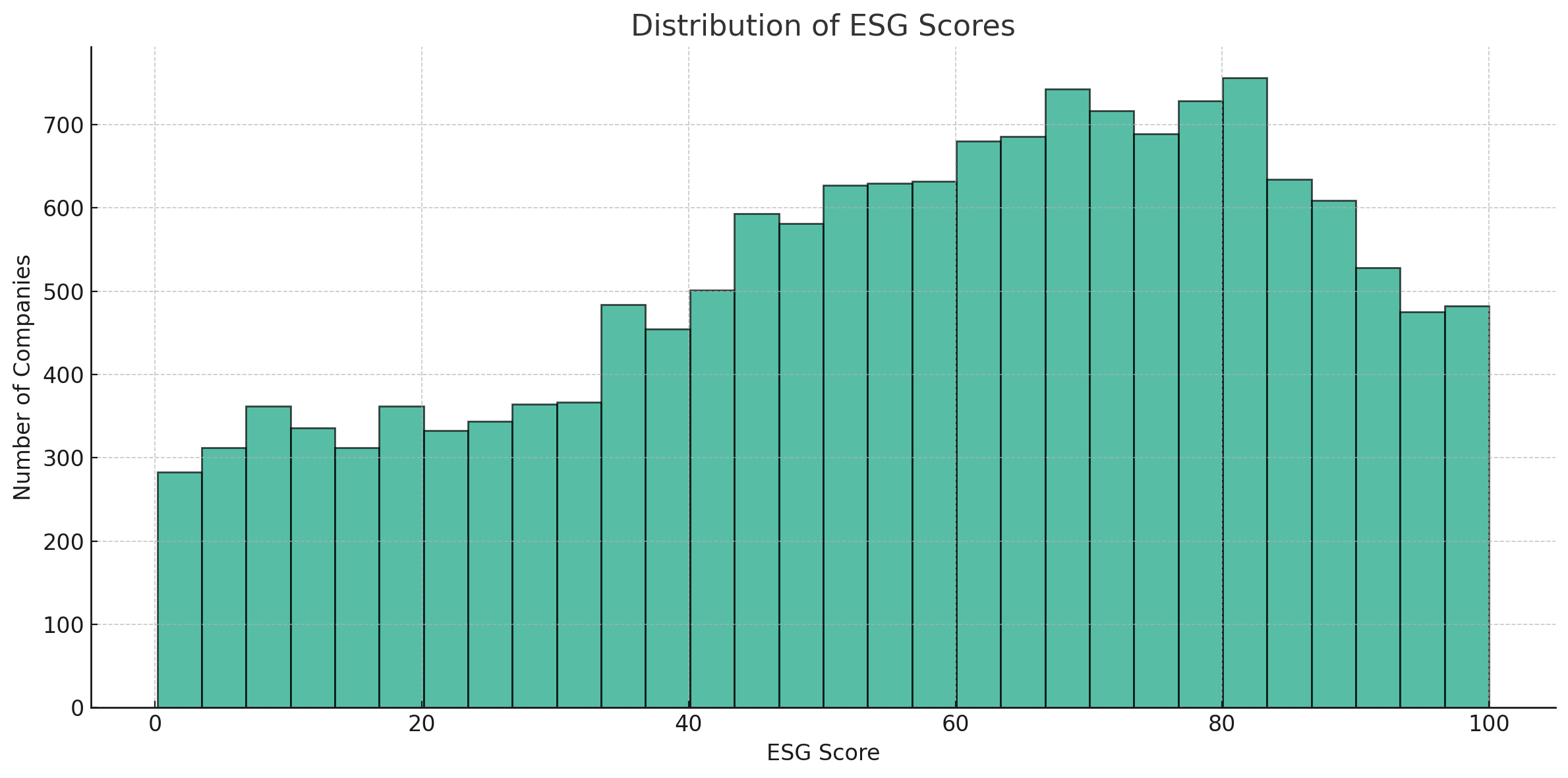}
\caption[Distribution of ESG Score]{Distribution of ESG Score \label{fig:esg_distribution}}
\end{figure}

\begin{figure}[H]
\centering
\includegraphics[width=0.5\textwidth]{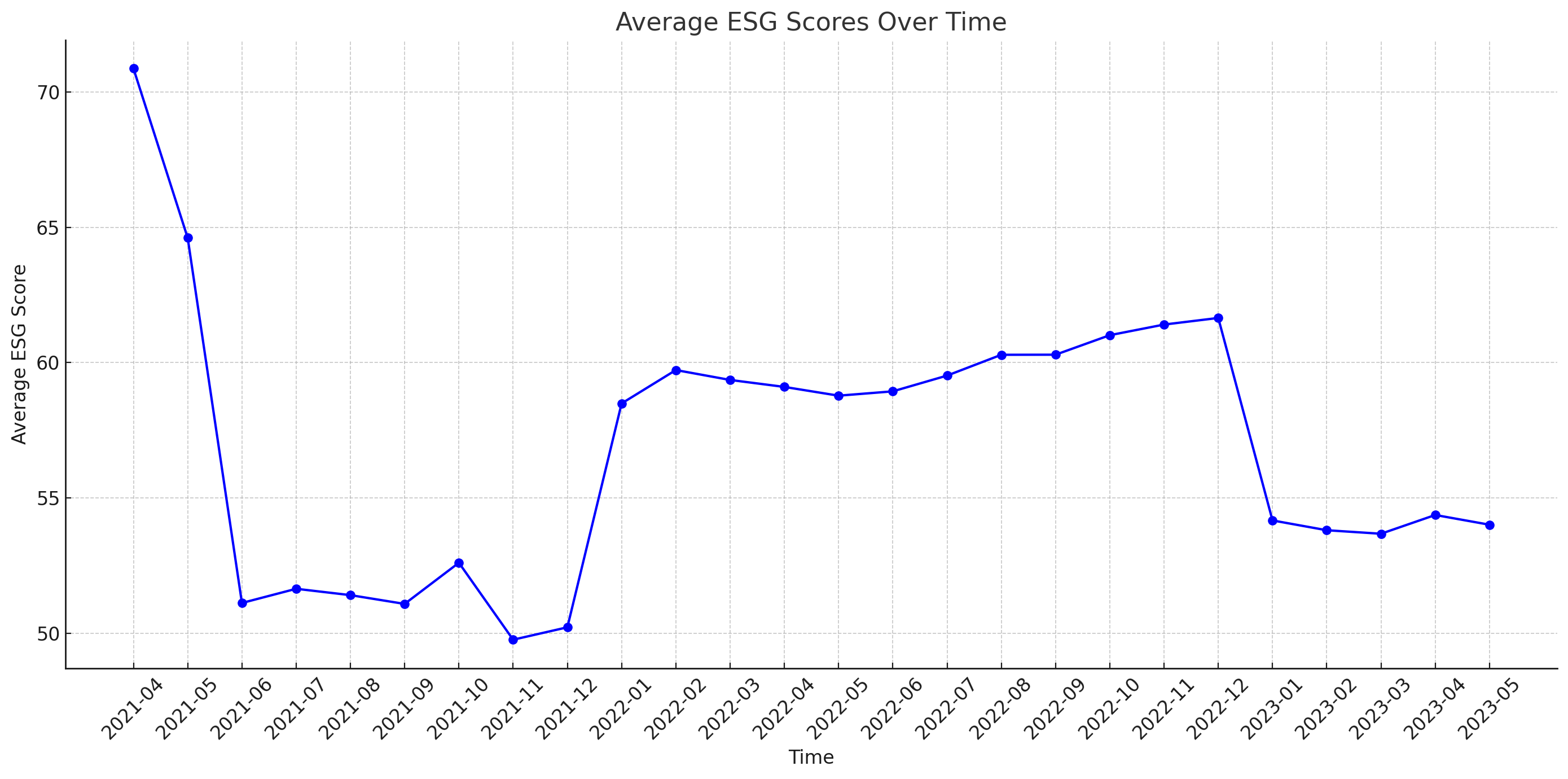}
\caption[Average ESG Scores Over Time]{Average ESG Scores Over Time \label{fig:avg_esg_scores}}
\end{figure}

In this section we present a description of data sourcing and their analysis, followed by the presentation of the approach adopted for selection of asset (stocks) for inclusion in the portfolio. The ESG data for firms of Indian market was obtained from the CSRHub online repository \cite{CSRHub}. This comprehensive data set includes monthly historical ESG records pertaining to a large number of firms, along with information identifying to the corresponding industries to which the firms' belong, offering detailed insight into the ESG practices adopted by these firms. In particular, our comprehensive exercise of extraction and collation of the data, resulted in a total of 133 unique industrial sectors and 1,392 unique firms, with the some of the key statistics resulting from the analysis of the data, are as enumerated below:
\begin{enumerate}[(A)]
\item The average ESG score across all firms was determined to be approximately $56.46$.
\item The average ESG score across all industries was approximately $54.91$.
\item Out of the total of 1,392 firms analyzed, 570 firms did not have recorded ESG scores.
\item The firm with the highest ESG score was FLEX-FOODS-LIMITED, with a perfect score of $100.0$.
\item The firm with the lowest ESG score was Tide-Water-Oil-India-Co-Ltd, with a score of $0$ (though might also indicate the absence of a recorded ESG score).
\end{enumerate}
The histogram of the distribution of ESG scores in Figure \ref{fig:esg_distribution}, from which we infer the following:  
\begin{enumerate}[(A)]
\item A significant number of firms have ESG scores in the range of approximately $40$ to $80$.
\item There's a peak around the score of $84$, followed by a decline. However there are several firms achieving perfect scores.
\item Fewer firms have very low ESG scores, with a slight rise in the number of firms having scores close to $0$ (though this may be attributed to the absence of data).
\end{enumerate}
The trends of average ESG scores are present in Figure \ref{fig:avg_esg_scores}, from which we surmise the following : 
\begin{enumerate}[(A)]
\item The ESG scores generally show an upward trend, indicating that, on average, firms are improving their ESG performance over time.
\item There are some fluctuations in the scores, but the overall trajectory seems positive.
\item The final dip maybe due to addition of more new data in 2023. 
\end{enumerate}
Finally the top 5 and bottom 5 industry ESG scores are given in Table \ref{tab:industry_top} and Table \ref{tab:industry_bot} respectively. 
\begin{table}[H]
\centering
\begin{tabularx}{\textwidth}{Xl}
\toprule
\textbf{Industry} & \textbf{Mean ESG Industry Score} \\
\midrule
Electrical Equipment Manufacturing, Architectural, Engineering, \& Related Services & $98.7440$ \\ \\
Electronic Equipment \& Instrumentation, Energy Equipment \& Services & $96.4656$ \\ \\
IT \& Network Services, Business Support Services & $96.3444$ \\ \\
Manufacturing, Electrical Equipment Manufacturing & $91.2384$ \\ \\
Heavy \& Civil Engineering Construction, Electronic Equipment \& Instrumentation, Machinery Manufacturing & 90.0116 \\
\bottomrule
\end{tabularx}
\caption{Top 5 industry ESG scores \label{tab:industry_top}}
\end{table}
\begin{table}[H]
\centering
\begin{tabularx}{\textwidth}{Xl}
\toprule
\textbf{Industry} & \textbf{Mean ESG Industry Score} \\
\midrule
Forestry \& Fishing & $1.085000$ \\ \\
Industrial Conglomerates, Architectural, Engineering, \& Related Services, Airport, Harbor Operations, \& Logistics & $3.950000$ \\ \\
Electronic Equipment \& Instrumentation, Household Appliance Manufacturing, Oil and Gas Extraction & $6.538333$ \\ \\
Personal Care Products & $8.467857$ \\ \\
Accounting, Tax Prep., \& Payroll Services & $16.147500$ \\
\bottomrule
\end{tabularx}
\caption{Bottom 5 industry ESG scores \label{tab:industry_bot}}
\end{table}
For the asset picking exercise, in this paper, we adopt two distinct methodologies for stock selection utilizing ESG scores, as enumerated below:
\begin{enumerate}[(A)]
\item Approach 1:
\begin{enumerate}[(i)]
\item Categorizing the firms according to their respective industries.
\item Computing the average ESG score for all firms within each industry for a specific timestamp.
\item From each industry group, the firm that exhibits the highest ESG score for that timestamp is identified. \item Only firms whose ESG score exceeds a threshold parameter $\zeta$ (contingent on the ESG value specified in the preceding step) are considered.
\end{enumerate}
\item Approach 2:
\begin{enumerate}[(i)]
\item Evaluation of each firm's ESG score in relation to the average score of its respective industry.
\item Select firms that outperform their industry average in terms of ESG scores by a given specified margin $\xi$.
\end{enumerate}
\end{enumerate}
The stocks selected based on the aforesaid approaches will be identified by the term ``stock universe'', and this will then be employed for subsequent portfolio construction.

\section{Methodology}
\label{Sec_Methodology}

In this section, we focus on the identification of stock pairs for the purpose of pairs trading, a strategy that is driven fundamentally via the notions of mean reversion and stationarity. While the former indicates a tendency of values (in a time series) reverting back to the average, the latter encapsulates situations where the statistical properties do not exhibit any temporal changes. For the purpose of this work, we adopt four tests, namely, unit root tests, Engle-Granger co-integration test, Hurst exponent and half-life test, all of which we briefly elaborate upon in the following discussion.

The unit root test that was chosen for this work is the Augmented Dickey-Fuller (ADF) test, whose purpose is to ascertain the presence of a unit root in the series, thereby implying the existence of non-stationarity in the series. If the ADF test statistic is more negative than the critical value (as obtained from Dickey and Fuller table), then it indicates stationarity of the series and in case it is less negative than the critical value, then it is suggestive of non-stationarity \cite{Chan2013}.

The co-integration tests due to Engle and Granger \cite{Engle1987} describes a comprehensive methodology to ascertain the co-integration of two time series, with the specific procedure being accomplished through a two time-step method. In the first step (of the procedure) an ordinary least square (OLS) approach is used to determine the coefficient of a linear relation assumed between the two times series. This is followed by the second step wherein the residual of the linear relation is tested for stationarity, which if it holds, implies the co-integration between the two time series.

In case of a time series for prices, the rate of price diffusivity can be quantitatively determined. If the (logarithmic) price series does not exhibit the geometric Brownian walk, then contingent on the Hurst exponent, one can conclude whether the series exhibits mean-reverting (anti-persistent), uncorrelated or trending (persistent) behaviour \cite{Chan2013}.

Finally, we dwell upon the key factor of mean-reversion half-life, which indicates the time taken for a series to revert to its mean level \cite{Chan2013}. Assuming a mean-reversion model, with a corresponding mean-reversion factor, one may conclude mean reversion (exponential decay), contingent on whether the mean reversion factor is positive (negative). In fact, the mean reversion speed is observed to be inversely proportional to the absolute value of mean-reversion factor.

\section{APO Based Trading Strategy}
\label{Sec_APO}

In this section we present our trading strategy based on the Absolute Price Oscillator (APO) \cite{Engle1987}, an approach used to identify potential trading opportunities contingent on the price momentum of assets. This strategy makes use of the difference observed between a fast and a slow Exponential Moving Average (EMA) to ascertain the entry and exit strategy for the trades. A rising and a falling APO indicates bullish and bearish momentum, respectively. The first step in our strategy involves the adoption of the following pair selection approach as enumerated below and summarized in Algorithm \ref{algo1}.
\begin{enumerate}[(A)]
\item Dimensionality Reduction:
This step involves processing the financial data to find a compact representation for each security. Dimensionality reduction is crucial when dealing with large datasets to reduce the computational load while still capturing the most significant features that contribute to the variability in the data. Techniques such as Principal Component Analysis (PCA) or Singular Value Decomposition (SVD) can be used to reduce the number of variables, with the former (PCA) being used in this work.
\item Clustering with OPTICS:
After reducing the dimensionality, the next step is to apply a clustering algorithm to group similar data points. OPTICS (Ordering Points To Identify the Clustering Structure) is a clustering algorithm \cite{optics} that can handle varying densities and is robust to outliers, making it suitable for financial datasets where such characteristics are common. This step will organize the securities into clusters based on their similarities in the reduced dimensional space.
\item Forming Pairs: Once the clusters are formed, the final step is to select all possible pair combinations within each cluster. Since securities within the same cluster are similar according to the chosen metrics and dimensions, they are likely candidates for pairs trading. This step involves generating pairs that will then be evaluated based on their trading potential, typically through metrics like co-integration, mean reversion, and other statistical measures relevant to pairs trading strategies \cite{Saramento2021}.
\end{enumerate}
\begin{algorithm}[H]
\caption{PCA-based clustering}\label{algo1}
\begin{algorithmic}[1]
\Procedure{Pairs Selection Framework}{}
\State \textbf{Dimensionality Reduction:}
\State Apply technique to find a compact representation for each security.
\State \textbf{Clustering with OPTICS:}
\State Apply OPTICS algorithm to identify clusters of similar data points.
\State \textbf{Forming Pairs:}
\State Select all possible pair combinations within each cluster.
\State \textbf{Output:} Clusters indicative of potential pairs for trading.
\EndProcedure
\end{algorithmic}
\end{algorithm}

In the second step, we introduce the APO spread trading strategy which capitalizes on the spread differences between two correlated assets. The algorithm computes a hedge ratio to form a spread series, which is then analyzed for trading signals based on the APO \cite{AlgoTrading}. This method provides a systematic approach to pairs trading, focusing on the principles of mean reversion. It leverages statistical methods to form a hedge position between two assets, aiming to profit from convergence in their price series. By continuously monitoring the APO spread, the strategy can adapt to market movements, making it a dynamic tool for traders. The specifics of the APO based approach adopted by us is enumerated in Algorithm \ref{algo2}.
\begin{algorithm}[H]
\caption{APO Spread Trading Strategy Algorithm}\label{algo2}
\begin{algorithmic}[1]
\State \textbf{Input:} Stock data for two assets $S_1$ and $S_2$, buy threshold, sell threshold
\State \textbf{Output:} Trade signals for pairs trading
\State
\Procedure{Compute Hedge Ratio}{data1, data2}
\State model $\gets$ perform OLS regression (data1, data2)
\State \Return model.params$[1]$
\EndProcedure
\State
\Procedure{Initialize}{fast, slow}
\State hedge$\_$ratio $\gets$ \Call{Compute Hedge Ratio}{$S_1$, $S_2$}
\State Calculate spread $\gets$ $S_1$$-$hedge$\_$ratio $\times$ $S_2$
\State fast$\_$ema $\gets$ \Call{EMA}{spread, fast}
\State slow$\_$ema $\gets$ \Call{EMA}{spread, slow}
\State apo$\_$spread $\gets$ fast$\_$ema$-$slow$\_$ema
\State position $\gets 0$
\EndProcedure
\State
\Procedure{Next}{buy threshold, sell threshold}
\State \textbf{if}{~apo$\_$spread $<$ buy$\_$threshold} \textbf{then}
\State \hspace{1cm} Execute Buy for $S_1$ and Sell Short for $S_2$
\State \textbf{else if}{apo$\_$spread $>$ sell$\_$threshold} \textbf{then}
\State \hspace{1cm} Execute Sell Short for $S_1$ and Buy for $S_2$
\State \textbf{end if}
\EndProcedure
\end{algorithmic}
\end{algorithm}

\section{Results}
\label{Sec_Results}

For the empirical analysis, our data set comprises a comprehensive selection of equities listed on the National Stock Exchange (NSE) \cite{NSEIndia} of India. In the initial data acquisition phase, we encountered the common challenge of data sparsity and non-uniformity, which necessitated a meticulous data cleaning process. We focused on ensuring that only stocks with a complete and consistent set of historical data were retained for further analysis, thereby eliminating any securities that could introduce bias or inaccuracy due to incomplete information. Post the data cleaning step, we applied the ESG filtering criterion and selected only those firms that demonstrated superior ESG performance relative to the industry average. The final data set, after rigorous cleaning and applying the ESG filter, consisted of a balanced panel of equities that are both representative of the market and reflective of high sustainability standards. This data set forms the backbone of our analysis, ensuring that our pairs trading strategy is both financially sound and ethically responsible. Finally, for the purpose of backtesting we used a commission rate of $0.1\%$.

The results for the training set are presented in Table \ref{tab:train_results} from where we see that pairs with high Sharpe Ratios and low drawdowns, like ``GESHIP and SOMANYCERA'' and ``BAJAJELEC and GSPL'', are generally more favorable as they suggest efficient risk-adjusted returns with lower declines. Pairs with high returns but also high drawdowns, such as ``CCL and SSWL'' and ``ADANIPORTS and ASHOKLEY'', might have provided good returns but at higher risk levels. Consistency between training and testing performance is crucial for validating any trading strategy, so it's important to compare these training results with out-of-sample testing results for a comprehensive assessment.
\begin{table}[H]
\centering
\begin{tabular}{lllll}
\toprule
     Pair1 &      Pair2 & Train Sharpe & Train Drawdown & Train Returns \\
\midrule
     GOKEX &   PETRONET &       0.5716 &         0.0000 &       18.5213 \\
JINDALPOLY &       RCOM &       0.6804 &         0.4312 &       24.9551 \\
    RADICO &  TATAELXSI &       0.6107 &         0.6194 &       20.4039 \\
 CENTRALBK &       MRPL &       0.6548 &         0.6935 &       15.4413 \\
JINDALPOLY &       MMTC &       0.6393 &         0.9320 &       23.6378 \\
   DCBBANK &  JSWENERGY &       0.5750 &         1.4613 &       20.4349 \\
 CENTRALBK & ENGINERSIN &       0.6737 &         1.6708 &        9.8932 \\
   DCBBANK &  SOUTHBANK &       0.5613 &         1.8784 &       21.6379 \\
     GOKEX &   GOCLCORP &       0.5239 &         1.9633 &       14.7174 \\
BAJAJ-AUTO &        ITC &       0.7972 &        12.7331 &       19.2181 \\
BAJAJHLDNG & MAHSCOOTER &       0.6238 &        14.1017 &       18.3774 \\
      GHCL &       SSWL &       0.7697 &        16.7510 &       13.4474 \\
   DCBBANK & ENGINERSIN &       0.5185 &         2.1551 &       18.4253 \\
 BAJAJELEC &     RADICO &       0.7588 &         2.1763 &       23.9585 \\
  HDFCBANK &  KOTAKBANK &       0.6458 &         2.3241 &        9.0718 \\
 BAJAJELEC &       GSPL &       0.9368 &         3.0490 &       21.8646 \\
BALRAMCHIN &       IIFL &       0.5144 &         3.3268 &       16.7721 \\
    GESHIP & SOMANYCERA &       1.1106 &         3.5970 &       19.3314 \\
   INDIANB &       MRPL &       0.5213 &        30.1832 &       19.3686 \\
ADANIPORTS &   ASHOKLEY &       0.5948 &        30.3840 &       13.4715 \\
BALRAMCHIN &     JKTYRE &       0.6734 &         4.1314 &       22.9861 \\
       HGS & MANAPPURAM &       0.5235 &         4.3003 &       23.6353 \\
       CCL &       SSWL &       0.5794 &        45.8747 &       13.4428 \\
INDUSINDBK & SHRIRAMFIN &       0.5808 &         6.7566 &       23.2811 \\
 CENTRALBK &        NCC &       0.5740 &         7.8748 &       12.0443 \\
\bottomrule
\end{tabular}
\caption{Training Results \label{tab:train_results}}
\end{table}

The wide range of Sharpe Ratios and Test Returns indicates that performance varies significantly across different pairs. From Table \ref{tab:test_results} we notice that pairs like ``RADICO and TATAELXSI'' and ``ADANIPORTS and ASHOKLEY'' show high Sharpe Ratios and positive returns, suggesting they performed well during the test period. On the other hand, pairs like ``CENTRALBK and ENGINERSIN'' and ``INDIANB and MRPL'' with negative Sharpe Ratios and returns indicate underperformance.
\begin{table}[H]
\centering
\begin{tabular}{lllll}
\toprule
     Pair1 &      Pair2 & Test Sharpe & Test Drawdown & Test Returns \\
\midrule
     GOKEX &   PETRONET &     -0.8136 &      194.0331 &  \textendash \\
JINDALPOLY &       RCOM &      0.1746 &        0.5390 &       2.3736 \\
    RADICO &  TATAELXSI &      2.3859 &        0.0000 &       7.4691 \\
 CENTRALBK &       MRPL &      0.0104 &        3.3137 &       1.1331 \\
JINDALPOLY &       MMTC &      0.1447 &        3.2516 &       1.7674 \\
   DCBBANK &  JSWENERGY &     -0.0275 &       15.0868 &      -2.9384 \\
 CENTRALBK & ENGINERSIN &     -6.6027 &        0.2840 &       0.1121 \\
   DCBBANK &  SOUTHBANK &     -0.7268 &        2.9508 &       0.4989 \\
     GOKEX &   GOCLCORP &     -0.9397 &      177.5813 &  \textendash \\
BAJAJ-AUTO &        ITC &     -0.8490 &       13.0009 &      -3.4101 \\
BAJAJHLDNG & MAHSCOOTER &     -0.3883 &       17.6147 &      -6.6273 \\
      GHCL &       SSWL &      1.4122 &        0.5998 &       7.2819 \\
   DCBBANK & ENGINERSIN &     -0.5181 &        0.2315 &       0.5511 \\
 BAJAJELEC &     RADICO &      0.8062 &        3.5841 &       6.4714 \\
  HDFCBANK &  KOTAKBANK &     -0.3050 &        5.9456 &       0.5962 \\
 BAJAJELEC &       GSPL &      2.5906 &        0.0000 &       7.0640 \\
BALRAMCHIN &       IIFL &      0.4547 &        0.3986 &       3.9414 \\
    GESHIP & SOMANYCERA &     -0.7121 &       38.4214 &     -11.6323 \\
   INDIANB &       MRPL &     -1.0069 &       17.9628 &      -3.5252 \\
ADANIPORTS &   ASHOKLEY &      5.5966 &        8.1725 &       7.4305 \\
BALRAMCHIN &     JKTYRE &     -0.3629 &       11.4830 &      -4.4709 \\
       HGS & MANAPPURAM &      0.2499 &        1.4495 &       0.5723 \\
       CCL &       SSWL &      5.7691 &        3.2371 &      11.1632 \\
INDUSINDBK & SHRIRAMFIN &      3.6843 &        0.5587 &       8.5831 \\
 CENTRALBK &        NCC &         nan &        0.0000 &       0.0000 \\
\bottomrule
\end{tabular}
\caption{Testing Results \label{tab:test_results}}
\end{table}

The cointegration values suggest how strongly the pairs are correlated in the long term. From Table \ref{tab:cointegration_hedge_revised} we infer that pairs like ``BALRAMCHIN and IIFL'' or ``DCBBANK and SOUTHBANK'' have a stronger cointegration, meaning these pairs are more likely to move together over time. Lower values, like in ``GOKEX and GOCLCORP'' or ``BAJAJHLDNG and MAHSCOOTER'' suggest weaker cointegration, indicating less predictability in the movement of these pairs together over time.
\begin{table}[H]
\centering
\begin{tabular}{llllll}
\toprule
     Pair1 &      Pair2 & Hedge Ratio & Cointegration &  Half-Life &   Cross \\
\midrule
     GOKEX &   PETRONET &     0.0266 &        0.0270 &   154.6689 & 17.0000 \\
JINDALPOLY &       RCOM &    -0.2093 &        0.3778 & -3108.3293 &  6.0000 \\
    RADICO &  TATAELXSI &     0.1881 &        0.7027 &   511.3864 &  3.0000 \\
 CENTRALBK &       MRPL &     0.6533 &        0.3883 &   165.3586 & 10.0000 \\
JINDALPOLY &       MMTC &     0.0082 &        0.2972 &   304.7383 & 12.0000 \\
   DCBBANK &  JSWENERGY &     0.5206 &        0.6992 &    90.1140 & 33.0000 \\
 CENTRALBK & ENGINERSIN &     0.7560 &        0.5032 &   175.0639 & 16.0000 \\
   DCBBANK &  SOUTHBANK &    -0.0049 &        0.7776 &  4966.0117 & 22.0000 \\
     GOKEX &   GOCLCORP &     0.0577 &        0.0055 &    62.4343 & 43.0000 \\
BAJAJ-AUTO &        ITC &     8.3471 &        0.3522 &   183.9925 & 21.0000 \\
BAJAJHLDNG & MAHSCOOTER &     0.7121 &        0.0035 &    80.9430 & 22.0000 \\
      GHCL &       SSWL &     2.4861 &        0.1172 &   337.4438 & 10.0000 \\
   DCBBANK & ENGINERSIN &     0.4730 &        0.5506 &   178.8774 & 15.0000 \\
 BAJAJELEC &     RADICO &     0.9440 &        0.0180 &   198.7051 & 17.0000 \\
  HDFCBANK &  KOTAKBANK &     0.7478 &        0.0009 &    81.9989 & 19.0000 \\
 BAJAJELEC &       GSPL &     1.8355 &        0.2210 &   311.5972 &  9.0000 \\
BALRAMCHIN &       IIFL &     0.2224 &        0.8221 &   289.8524 & 13.0000 \\
    GESHIP & SOMANYCERA &     0.1787 &        0.0192 &   131.8533 & 12.0000 \\
   INDIANB &       MRPL &     2.6927 &        0.0561 &   188.8532 & 22.0000 \\
ADANIPORTS &   ASHOKLEY &     2.2631 &        0.3997 &   442.6782 &  7.0000 \\
BALRAMCHIN &     JKTYRE &     0.4465 &        0.8090 &   367.8877 &  5.0000 \\
       HGS & MANAPPURAM &     1.2358 &        0.1436 &   649.1161 &  7.0000 \\
       CCL &       SSWL &     2.8661 &        0.6521 &   440.5626 & 10.0000 \\
INDUSINDBK & SHRIRAMFIN &     1.8646 &        0.0073 &    44.4288 & 41.0000 \\
 CENTRALBK &        NCC &     0.3387 &        0.6469 &   353.2915 &  5.0000 \\
\bottomrule
\end{tabular}
\caption{Cointegration and Hedge Ratio without Mean Reversion \label{tab:cointegration_hedge_revised}}
\end{table}

\begin{figure}[H]
\centering
\includegraphics[width=0.4\textwidth]{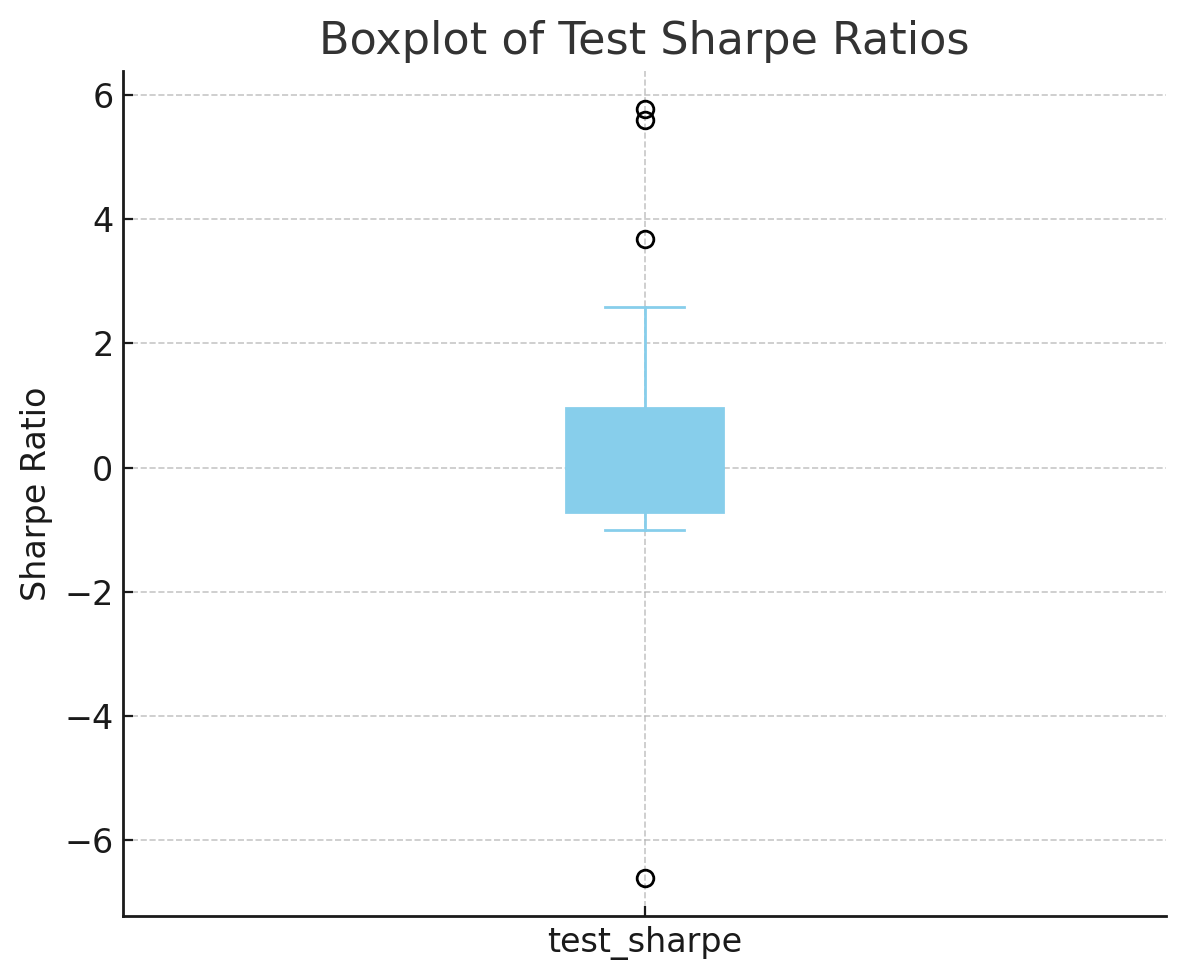}
\caption[Boxplot of Sharpe Ratios]{Boxplot of Sharpe Ratios \label{fig:bp_sharpe}}
\end{figure}

\begin{figure}[H]
\centering
\includegraphics[width=0.4\textwidth]{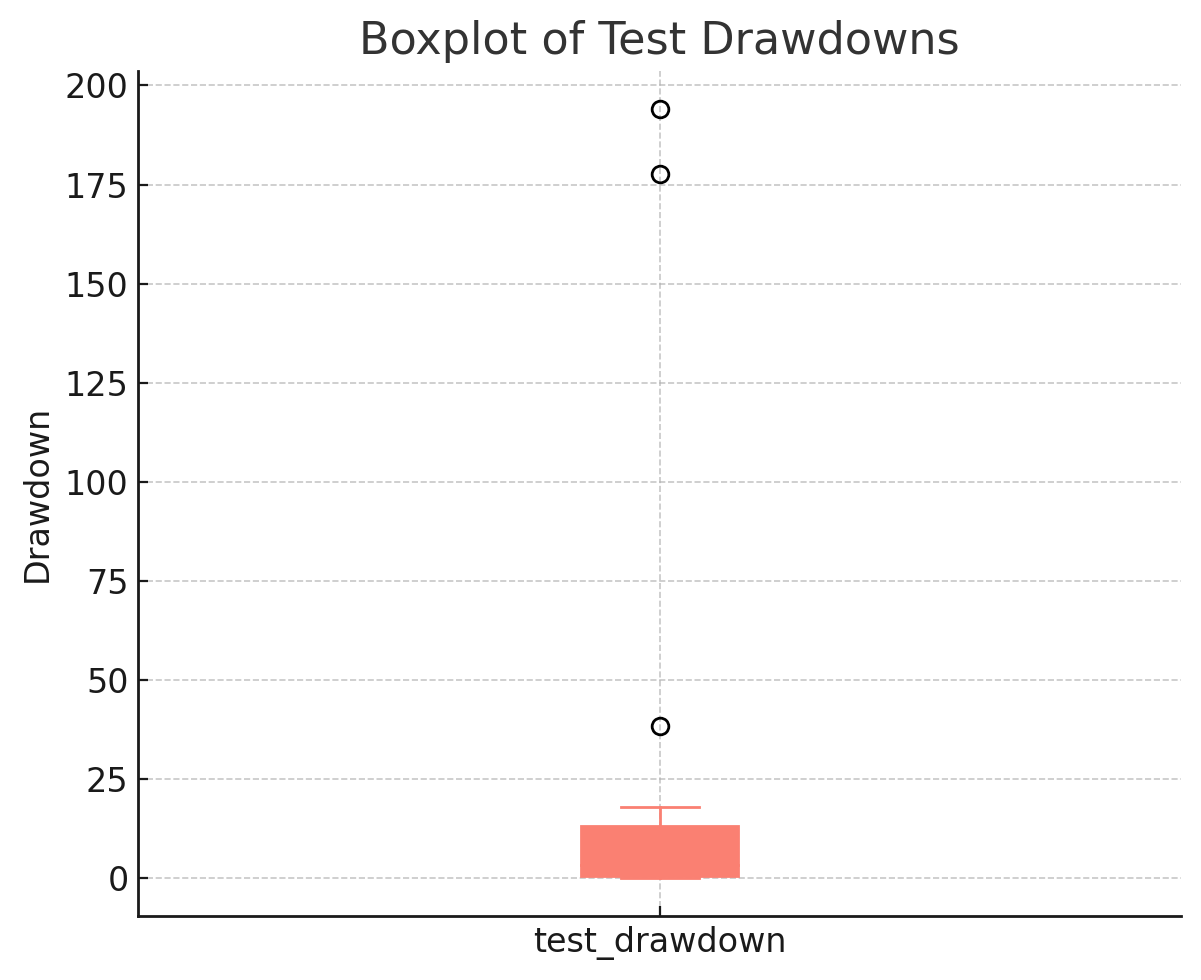}
\caption[Boxplot of Drawdown]{Boxplot of Drawdown \label{fig:bp_drawdown}}
\end{figure}

\begin{figure}[H]
\centering
\includegraphics[width=0.4\textwidth]{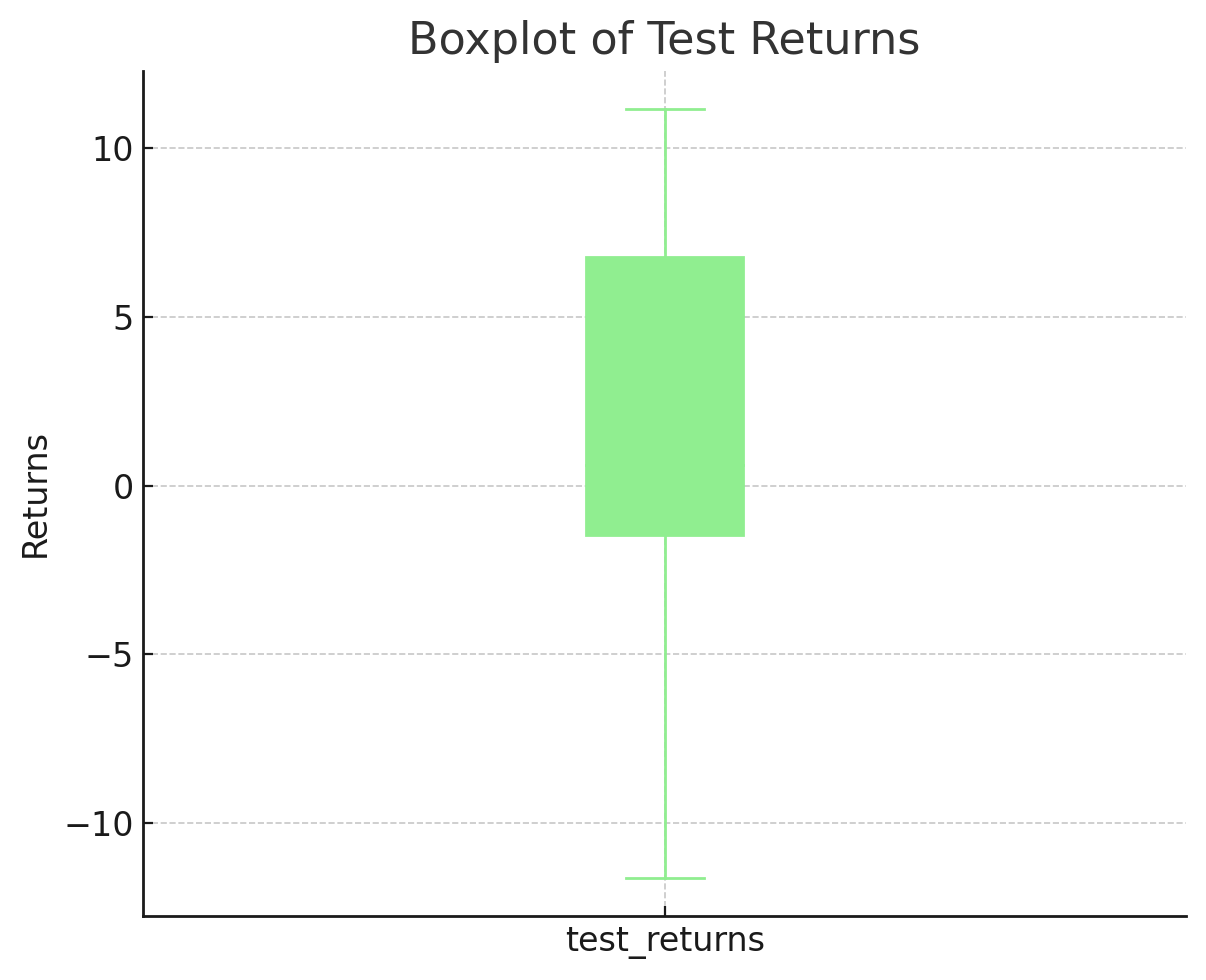}
\caption[Boxplot of Returns]{Boxplot of Returns \label{fig:bp_returns}}
\end{figure}

Finally, We can draw various conclusions about the data by plotting the box plots. The box plot for Sharpe ratios in Figure \ref{fig:bp_sharpe} indicates a median near zero, with a wide interquartile range and outliers in both directions, signifying a mix of pairs with both high and low risk-adjusted returns. The drawdown box plot in Figure \ref{fig:bp_drawdown} shows a low median with most data clustered in a narrow range, suggesting consistent potential drawdown across pairs, yet with some notable high-drawdown outliers. The returns box plot in Figure \ref{fig:bp_returns} displays a concentration of pairs with modest median returns, a tight interquartile range, and a few exceptional outliers.

\section{Conclusion and Future Directions}
\label{Sec_Conclusion}

This paper introduces a pairs trading algorithm combining ESG measures for sustainable asset picking. Quantitative techniques, including, cointegration, mean reversion and dimensionality reduction are combined with the APO spread trading approach to algorithmically identify and trade pairs. Although backtests show differences between train and test results, some top pairs display strong signals, integration and positive risk-adjusted gains over both (train and test) periods, indicating feasibility, in terms real-life trading. Overall this study connects ethical investing and statistical arbitrage, providing a framework positioned to traverse markets, while upholding sustainability values.

Future work would involve application of this strategy to high-frequency data in order to uncover more trading opportunities and understand short term price moves, despite increased complexity. Additional dimensionality reduction methods like t-SNE could reveal new data patterns missed by PCA. Exploring clustering algorithms beyond OPTICS may provide fresh insights and alternative pairs. Machine learning approaches could dynamically determine entry and exit points, thereby adapting the strategy to evolving markets. The model could also integrate other performance factors like macroeconomic and sentiment indicators or more nuanced ESG metrics. In summary, integrating high-frequency data, new dimensionality and clustering techniques, and advanced machine learning promises an adaptive, robust strategy aligning of financial and sustainability objectives.

\section*{Declaration of Interests}

The authors declare that they have no known competing financial interests or personal relationships that could have appeared to influence the work reported in this paper.

\bibliographystyle{plain}
\bibliography{bib.bib}

\begin{thebibliography}{10}

\bibitem{optics}
Mihael Ankerst, Markus~M Breunig, Hans-Peter Kriegel, and J{\"o}rg Sander.
\newblock {OPTICS}: Ordering points to identify the clustering structure.
\newblock {\em ACM Sigmod record}, 28(2):49--60, 1999.

\bibitem{Chan2013}
Ernie Chan.
\newblock {\em Algorithmic trading: winning strategies and their rationale},
  volume 625.
\newblock John Wiley \& Sons, 2013.

\bibitem{CSRHub}
{CSRHub}.
\newblock {CSRH}ub - corporate social responsibility and sustainability
  ratings.
\newblock \url{https://www.csrhub.com/}, 2023.
\newblock Accessed: 2023-10-04.

\bibitem{AlgoTrading}
Sebastien Donadio and Sourav Ghosh.
\newblock {\em Learn Algorithmic Trading: Build and deploy algorithmic trading
  systems and strategies using Python and advanced data analysis}.
\newblock Packt Publishing Ltd, 2019.

\bibitem{Engle1987}
Robert~F Engle and Clive~WJ Granger.
\newblock Co-integration and error correction: representation, estimation, and
  testing.
\newblock {\em Econometrica: Journal of the Econometric Society}, pages
  251--276, 1987.

\bibitem{EU_ETS}
{European Commission}.
\newblock {EU} {E}missions {T}rading {S}ystem ({EU ETS}).
\newblock
  \url{https://climate.ec.europa.eu/eu-action/eu-emissions-trading-system-eu-ets_en}.
\newblock Accessed: 2024-10-04.

\bibitem{Hellmich2021}
Martin Hellmich and Rudiger Kiesel.
\newblock {\em Carbon Finance: {A} Risk Management View}.
\newblock World Scientific, 2021.

\bibitem{ISDA}
{International Swaps and Derivatives Association}.
\newblock Role of derivatives in carbon markets.
\newblock
  \url{https://www.isda.org/a/soigE/Role-of-Derivatives-in-Carbon-Markets.pdf}.
\newblock Accessed: 2024-01-24.

\bibitem{NSEIndia}
{National Stock Exchange of India}.
\newblock {NSE} - national stock exchange of india ltd.
\newblock \url{https://www.nseindia.com/}, 2023.
\newblock Accessed: 2023-10-11.

\bibitem{Roncalli2022}
Thierry Roncalli.
\newblock Handbook of {S}ustainable {F}inance.
\newblock {\em Available at SSRN}, 2022.

\bibitem{Saramento2021}
Sim{\~a}o~Moraes Sarmento and Nuno Horta.
\newblock {\em A Machine Learning based Pairs Trading Investment Strategy}.
\newblock Springer, 2021.

\bibitem{unpri}
{UN Principles for Responsible Investment}.
\newblock Climate change and the just transition: {A} guide for investor
  action.
\newblock \url{https://www.unpri.org/download?ac=9452}, 2018.
\newblock Accessed: 2023-10-04.

\bibitem{wbcsd}
{World Business Council for Sustainable Development}.
\newblock {R}eporting {M}atters 2020.
\newblock \url{https://www.wbcsd.org/contentwbc/download/10460/156310/1}, 2020.
\newblock Accessed: 2023-10-04.

\bibitem{wri}
{World Resources Institute}.
\newblock {S}tate of {C}limate {A}ction 2022.
\newblock
  \url{https://www.wri.org/webform/download_publication?source_entity_type=node&source_entity_id=102515},
  2022.
\newblock Accessed: 2023-10-04.

\end{thebibliography}

\end{document}